\documentstyle[epsfig,longtable]{aipproc}
\begin{document}
\title{QCD Pomeron at Linear Colliders\footnote{work supported by the
Graduiertenkolleg "Theoretische Elementarteilchenphysik".}}
\author{Cong-Feng Qiao}
\address{II Institut f\"ur  Theoretische Physik, Universit\"at Hamburg\\
Luruper Chaussee 149, D-22761 Hamburg, Germany.}
\maketitle
\vspace {-6mm}
\begin{abstract}
Recent developments in theory on the calculation of $\gamma^* \gamma^*$
reaction at high energies, in the aim of detecting the BFKL Pomeron
signals, are briefly introduced. The importance of the NLO QCD
corrections to the Photon Impact Factor in the game is highlighted.
\end{abstract}
\section*{Introduction}
The experimental test of the BFKL Pomeron \cite{klf} is generally
considered to be an important task in strong interaction(QCD) physics. In
past years, it has been studied in the small-x region at HERA intensively
in many respects \cite{nrw}. However, because of the non-perturbative
effects on both proton and final state hadronization, there still have no
definite conclusions been made.

Following the pioneering application of the BFKL formalism to high energy
$\gamma^* \gamma^*$ interactions at Linear Colliders by Bartels
{\it et al.} \cite{b1} and Brodsky {\it et al.} \cite{lotter}, recently
much interest has been given to this kind of reaction \cite{motyka}.
This process describes the scattering of two small-size projectiles, and
its high energy(not too much high) behavior is expected to be
described by the BFKL Pomeron. Moreover, in $\gamma^* \gamma^*$ scattering
process the detection of the BFKL signal has several advantages comparing
to that in the deep inelastic scattering or hadron-hadron interactions.
One of them is that at Linear Colliders one can carefully calibrates the
scattering angles and energies of the outgoing leptons in order to choose
both interacting photons have the same or similar virtualities, and then
may get rid of or suppress the effect coming from the DGLAP evolution. 

\section*{The $\gamma^*\,\gamma^*$ Scattering in BFKL Formalism}

Before the advent of the field theory for strong interaction, the QCD, it
is well known that the Regge theory performed excellently in describing
the scattering of particles at high energies with strong interaction. 
With getting more and more confidence in QCD being as the correct field
theory for strong interaction, people began to investigate the QCD in
various aspects in details, and try to compare its predictions with what
obtained in Regge theory as well. The BFKL Pomeron, developed more than
twenty years ago, is one of the counterparts of those in Regge theory, 

\vskip 3mm
\begin{figure}[thb]
  \begin{center}
\epsfig{file=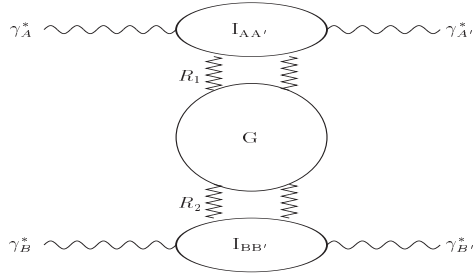,bbllx=220pt,bblly=280pt,bburx=380pt, 
bbury=420pt,width=4cm,height=2.5cm,clip=0}
  \end{center}
  \caption{Schematic representation of the process $\gamma^*_A\,
\gamma^*_B \rightarrow  \gamma^*_{A'}\, \gamma^*_{B'} $.}
  \label{graphs2}
\end{figure}

\vskip -2mm
In Regge limit, to Leading Order(LO) in Log$(\hat{s})$, the interaction of
$e^+\, e^- \rightarrow \, e^+\, e^- \, + X$ process can be formatted as 
a gluon ladder with the idea of Reggeization, where the $\sqrt{\hat{s}}$
is the center-of-mass energy of $\gamma^* \gamma^*$ system. In
this case the virtualities of the two photons would be large enough for
the pQCD to be used reliably, but not too large to ensure it is still in
the small-x domain, that is $Q_1^2,\, Q_2^2 \gg \Lambda^2_{\rm QCD}$ and
$x_1 = \frac{Q_1^2}{2 q_1\cdot{k_2}}\, $, $ x_2 = \frac{Q_2^2}{2
q_2\cdot{k_1}}\, \ll 1$. For instance, the process of $\gamma^*_A \,
\gamma^*_B \rightarrow \gamma^*_{A'} \, \gamma^*_{B'}$ can be
schematically shown as Figure 1, where the upper and lower blobs represent
the Pomeron Photon interactions, the Photon impact Factor(FIP) in
configuration space,  and to LO can be well understood in the language of
Photon wave function \cite{muller}, while the blob G in the middle
represents the Green's function of the two interacting Reggeized gluons. 

\section*{Comparison with the Experimental Data}

\vskip -1mm
\begin{figure}[htb]
\epsfig{file=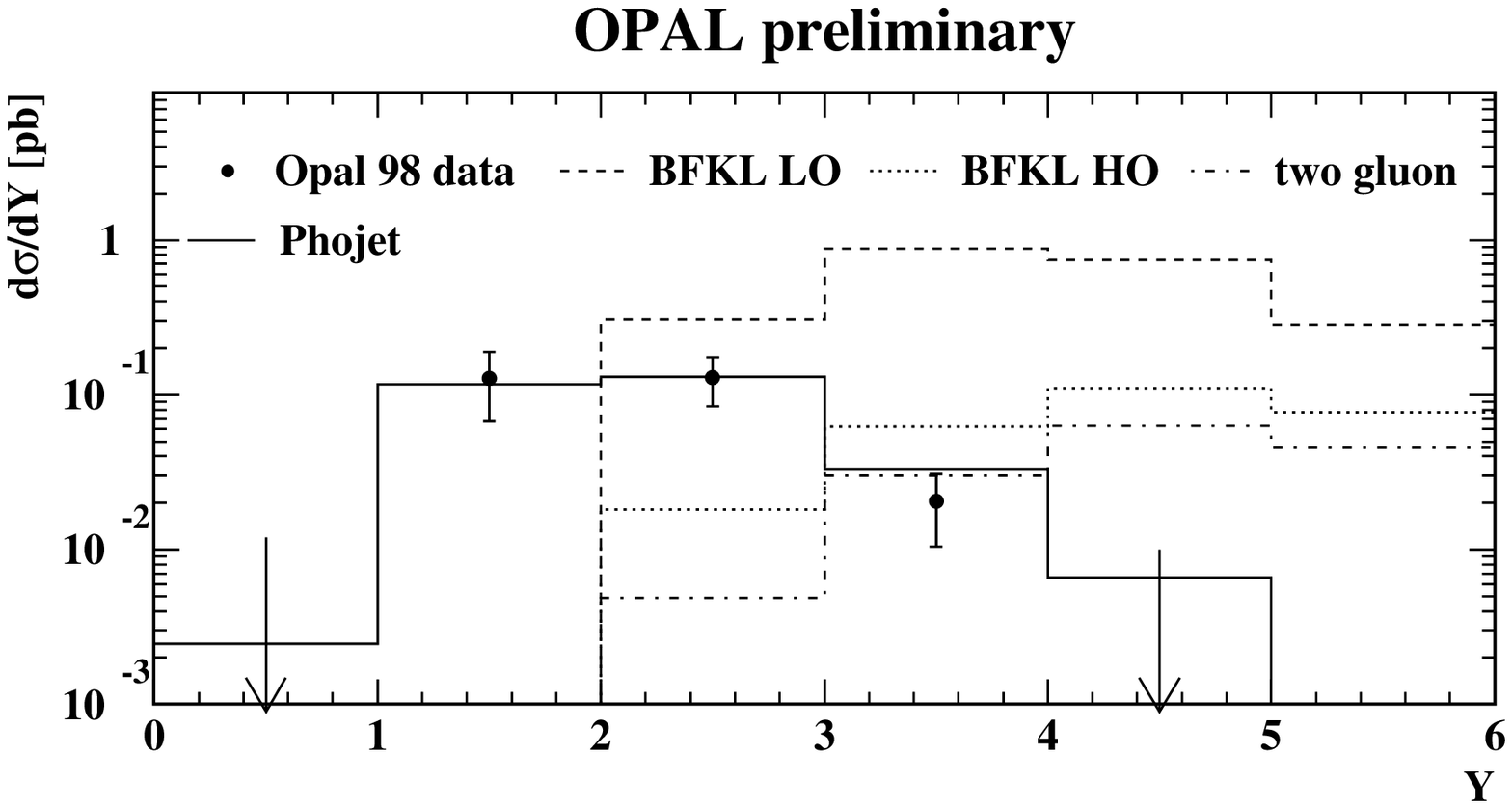,width=0.476\textwidth} 
\epsfig{file=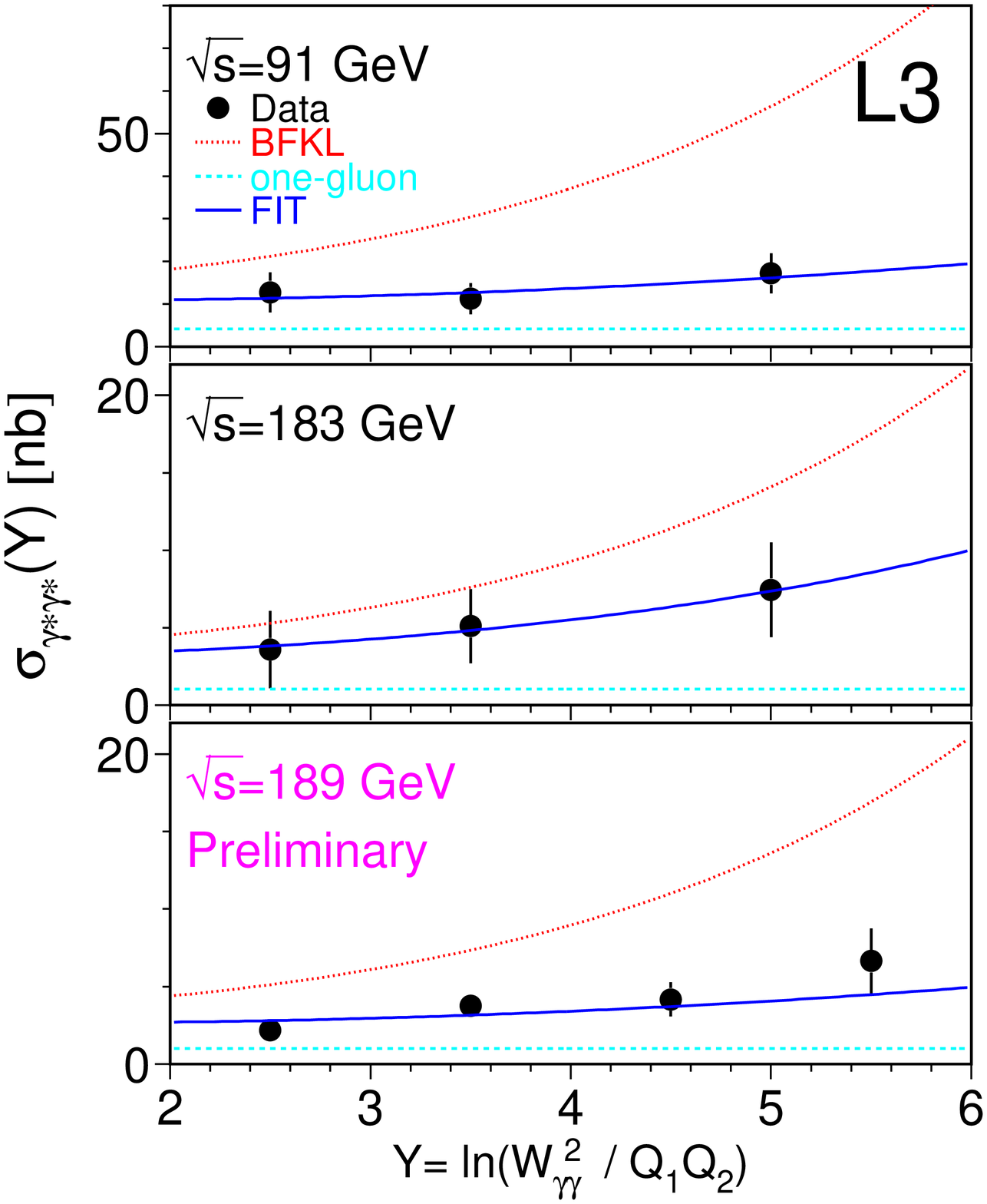,width=0.476\textwidth,height=4cm} 
\caption{Cross-section of the the process $e^+\, e^- \rightarrow
        e^+\, e^-\, \gamma^*\,\gamma^* -> e^+\, e^- + X$ as function of
        $Y \approx \rm{Log}\frac{\hat{s}}{s_0}$.}
  \label{opall3}
\end{figure}
\vspace {-3mm}
So far the LO calculations have been compared to the LEP data
(both OPAL and L3) \cite{lep}. Although the next-to-leading order (NLO)
corrections to the BFKL equation have been obtained two year ago
\cite{fl}, a complete comparison of the NLO BFKL calculations with the
measurements and making predictions for Future Linear Colliders (FLC) is
impossible since PIF is still unknown to the same order of accuracy.

In the LO comparisons, the data from both OPAL and L3 groups
lie above the two gluon exchange curve, the Born approximation,
but below the BFKL prediction, as shown in Figure \ref{opall3} \cite{rem},
where the $W_{\gamma\gamma}^2 = \hat{s}$ and $\sqrt{Q_1^2 Q_2^2} = s_0$.
Because we know the NLO corrections to BFKL equation is very large, the
comparison to go beyond the LO is necessary. By taking the BLM
scale setting scheme \cite{brodsky}, Brodsky {\it et al.} first attempted
to use the NLO BFKL equation to make the prediction on $\gamma^* \,
\gamma^*$ scattering \cite{kim}, and their result is encouraging, fits to
date pretty well. However, the question still remains that the NLO PIF is
unknown, which for sure would cast some doubts on the result.

The reason why radiative corrections to the PIF are important is not only
because the corrections may large, but also for the consistency reason of
the BFKL formalism. One of the major uncertainties in doing theoretical
calculations at LO is the energy scale $s_0$, which is not determined at
LO and hence the choice of it with some arbitrariness, while its
influence to the LO predictions is considerably large \cite{b1}. However,
the problem in $s_0$ is much less serious in the full NLO calculation, or
even can be fixed. Another source of uncertainty in the LO calculation is
from the renormalization scale $\mu^2$, which also incurs large errors in
the calculation \cite{lotter}. As well, the scale dependence will be
reduced at the NLO level.

\section*{The Calculation of QCD Corrections to Photon Impact Factor}

\begin{figure}[thb]
\begin{center}
\epsfig{file=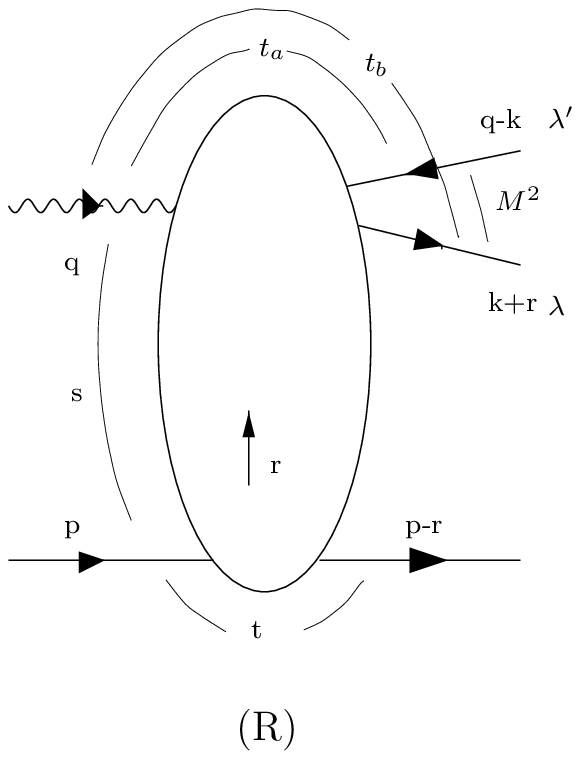,bbllx=0pt,bblly=550pt,bburx=135pt,
bbury=730pt,width=4cm,height=3.5cm,clip=0}
\epsfig{file=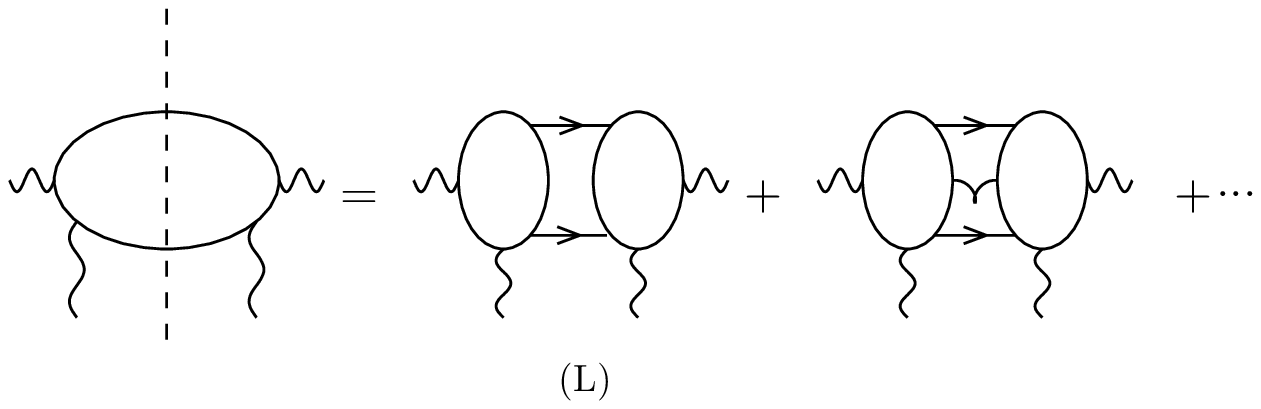,bbllx=455pt,bblly=650pt,bburx=635pt,
bbury=740pt,width=4cm,height=3.5cm,clip=0}
\end{center}
\caption{(L)Contributions to the photon impact factor; (R)Kinematics of
the process $\gamma^* + q \to q\bar{q} + q$.}
\label{kinematics}
\end{figure}
\vspace {-2mm}
The calculations of the NLO corrections to the PIF are still on the way
\cite{bgq}\cite{kotsky}. As shown in Figure \ref{kinematics}(L), the PIF
in forward case can be obtained from the energy discontinuity of the
amplitude $\gamma^* + \mbox{\textsl{Reggeon}} \to \gamma^* +
\mbox{\textsl{Reggeon}}$.  At LO in $\alpha_s$ it is simply the square of
the scattering amplitude $\gamma^*$ and a Reggeized gluon. In NLO
calculation, many possibilities of corrections should be considered.  The
task of calculating the NLO corrections to the PIF can be organized into
three steps, (i) the calculation of the NLO corrections to the $\gamma^* +
\mbox{\textsl{Reggeon}} \to q\bar{q}$ vertex, (ii) the vertex $\gamma^*
\to q\bar{q}g$ in LO, and (iii) the integration over the phase
space of the intermediate states.

Up to now, the first of those three steps has been finished \cite{bgq}. To
get the NLO Reggeon-Photon Coupling, for simplicity the process $\gamma^*
+ q \to q\bar{q} + q$ in Regge limit, that is $t,\; Q^2,\; t_a,\; t_b,\;
M^2 \; \ll \hat{s}$, has been considered, as shown in Figure
\ref{kinematics}(R). The results obtained possess natures as expected,
like  $t_a$ $t_b$ symmetry, etc. Because of complexity of the results in
NLO Reggeon-Photon vertex, the following two steps left in calculating the
NLO PIF are obviously not an easy task. 

\section*{Summary and Conclusions}

We have briefly reviewed the theoretical status in the research of finding
QCD Pomeron signals in $\gamma^* \, \gamma^*$ reaction. The current 
comparisons of theoretical calculations with the LEP data are encouraging,
but still pre-mature. A measurement of the reaction $e^+e^- \to e^+e^- +
X$ by tagging the outgoing leptons at FLC may provide an excellent test
of the very important QCD prediction, the BFKL Pomeron.

Again, we would like to point out that the LEP date for the total
$\gamma^* \, \gamma^*$ cross section clearly indicate that the LO
BFKL and two-gluon model are not sufficient to describe the data. To make
a consistent NLO prediction for FLC or comparison with the LEP data, the
still unknown PIF with NLO QCD corrections is an inevitably ingredient.

\vspace {-2mm}
\acknowledgments{I would like to express his gratitude to J. Bartels and
S. Gieseke for reading the manuscript and giving suggestions.}

\end{document}